\documentclass[a4paper, 10pt, conference]{ieeeconf}      

\IEEEoverridecommandlockouts                              

\usepackage{url}
\usepackage{textgreek}
\usepackage{array}
\usepackage{amsmath}
\usepackage{graphicx}
\usepackage{hyperref}

\usepackage{tabu}
\usepackage{siunitx}
\newcolumntype{D}{S[
 tight-spacing=true,round-mode=places,
                          round-precision=4,
                          table-format =  1.4]}

\usepackage{pgfplots}
\usetikzlibrary{shapes,shapes.geometric,arrows,fit,calc,positioning,automata,}
\pgfplotsset{compat=1.5}

\usepackage{flushend}

\usepackage{color}
\definecolor{dkgreen}{rgb}{0,0.6,0}
\definecolor{gray}{rgb}{0.5,0.5,0.5}
\definecolor{mauve}{rgb}{0.58,0,0.82}
\definecolor{orange}{gray}{0.1}
\definecolor{blues1}{gray}{0.75}
\definecolor{blues2}{gray}{0.7}
\definecolor{blues3}{gray}{0.65}
\definecolor{blues4}{gray}{0.5}
\definecolor{blues5}{gray}{0.4}
\usepackage{listings}
\title{\LARGE \bf
Systematic Evaluation of Sandboxed Software Deployment for Real-time Software on the Example of a Self-Driving Heavy Vehicle
}

\author{Philip Masek$^{1}$, Magnus Thulin$^{1}$, Hugo Andrade$^{1}$, Christian Berger$^2$, and Ola Benderius$^{3}$
\thanks{$^{1}$Department of Computer Science and Engineering, Chalmers University of Technology, G\"oteborg
        {\tt\small sica@chalmers.se}}%
\thanks{$^{2}$Department of Computer Science and Engineering, University of Gothenburg, G\"oteborg
        {\tt\small christian.berger@gu.se}}%
\thanks{$^{1}$Department of Applied Mechanics, Chalmers University of Technology, G\"oteborg
        {\tt\small ola.benderius@chalmers.se}}%
\thanks{~}
\thanks{We are grateful to Chalmers University of Technology to support the research in preparation for the 2016 GCDC competition. The Chalmers Revere laboratory is supported by Chalmers University of Technology, AB Volvo, Volvo Cars, and Region V\"astra G\"otaland.}
}

\begin{document}

\maketitle
\thispagestyle{empty}
\pagestyle{empty}

\begin{abstract}

Companies developing and maintaining \textit{software-only} products like 
web shops aim for establishing persistent links to their software running 
in the field. Monitoring data from real usage scenarios allows for a 
number of improvements in the software life-cycle, such as quick 
identification and solution of issues, and elicitation of requirements from 
previously unexpected usage. While the processes of continuous 
integration, continuous deployment, and continuous experimentation 
using sandboxing technologies are becoming well established in said 
software-only products, adopting similar practices for the automotive 
domain is more complex mainly due to real-time and safety constraints. 
In this paper, we systematically evaluate sandboxed software deployment 
in the context of a self-driving heavy vehicle that participated in the 2016 
Grand Cooperative Driving Challenge (GCDC) in The Netherlands. We 
measured the system's scheduling precision after deploying applications 
in four different execution environments. Our results indicate that there is 
no significant difference in performance and overhead when sandboxed 
environments are used compared to natively deployed software. Thus, 
recent trends in software architecting, packaging, and maintenance using 
\textit{microservices} encapsulated in sandboxes will help to realize similar 
software and system engineering for cyber-physical systems.
\end{abstract}

\section{Introduction} 

Companies that produce and maintain \textit{software-only} products,
i.e.~products that are only constituted with software and are
pre-dominantly executed in cloud-environments like web shops
or cloud-based applications, have started to strive for achieving
shorter product deployment cycles. Their goal is to tighten software
integration and software deployment up to their customers. When
these processes are reliably established, companies are able to
closely monitor their products in real usage scenarios from their
customers and can collect highly relevant data thereabouts. This
data is of essential business interest to better understand the
products in the field, maintain them in case of issues, and to
gradually introduce new features and observe their adoption.

For example, companies like Facebook and Netflix have established
mechanisms to enable continuous deployment that would allow
software updates as often as hundreds of times a day \cite{Savor2016}. 
These companies ultimately aim for a software engineering process
that integrates near real-time product feedback data as crucial method for
continuous business and product innovation.

A key-enabling technology for such software-only products is
resource isolation that is strictly separating system resources like
CPU time, network devices, or inter-process communication
resources; today's leading environments are the tool suite
Docker \cite{WWW_docker,Mer14} or Google's lmctfy
(from \textit{let me contain that for you}) \cite{WWW_lmctfy} that encapsulates
cloud-based applications. The key concept is to package all
relevant software product dependencies in self-contained
bundles that can be easily deployed, traced, archived, and even
safely rolled back in case of issues.

This study is an extension of a master thesis presented in \cite{masek}, where
further details are presented and available.

\begin{figure}[t]
  \begin{center}
	\includegraphics[width=.8\linewidth,trim=0cm 0cm 0cm 0cm]{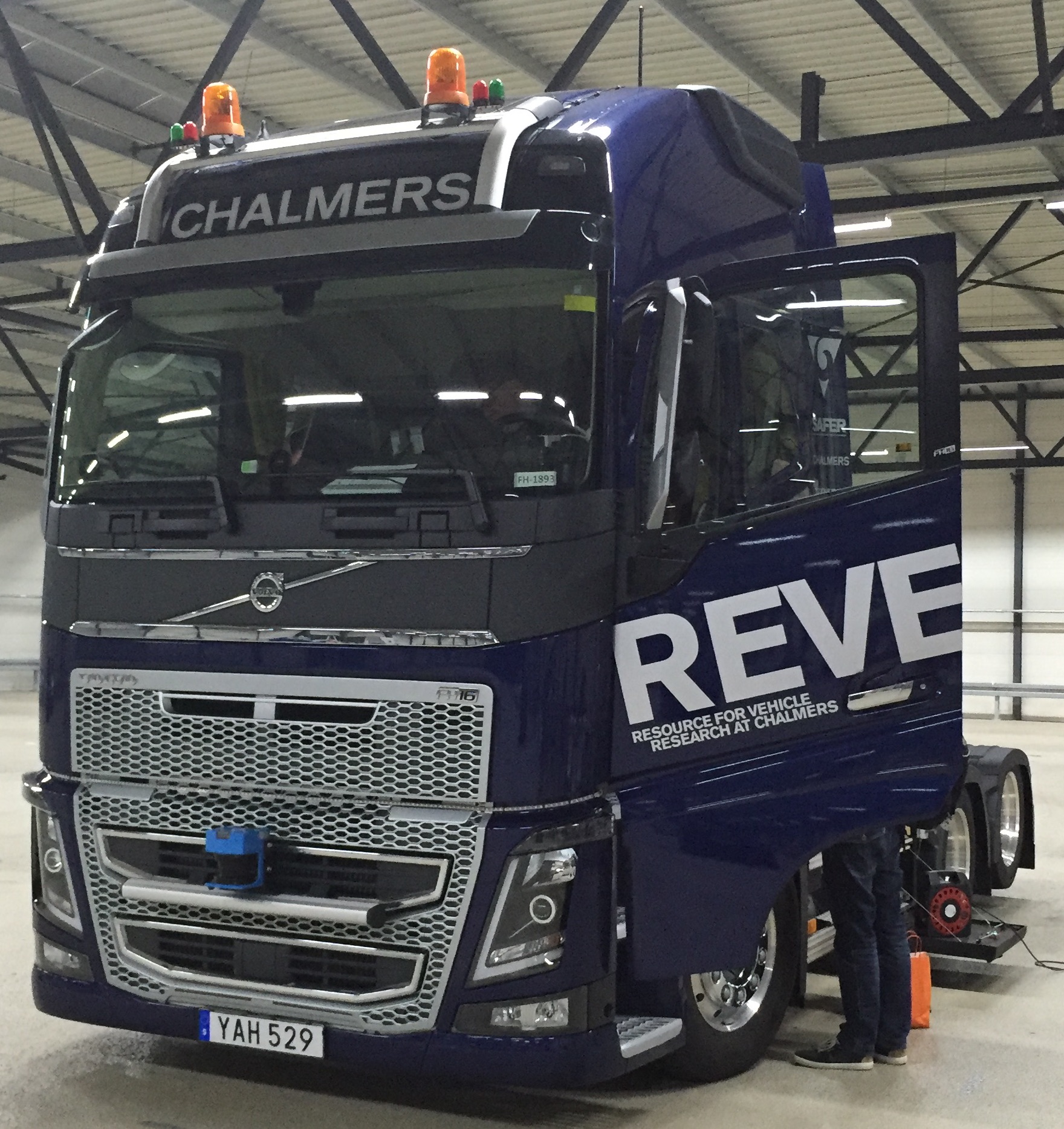}
        \caption{Volvo FH16 truck from Chalmers Resource for Vehicle Research (Revere) laboratory that participated in the 2016 Grand Cooperative Driving Challenge in Helmond, NL.}
        \label{fig:Truck}
  \end{center}  
\end{figure}

\subsection{Problem Domain \& Motivation}

Continuously developing, integrating, and deploying is challenging with 
software-only systems, i.e., systems that are not dependent on 
their surrounding physical environments. However, these tasks 
become even more challenging in the domain of self-driving 
vehicles, where applications are safety-critical and most of the 
times, even the underlying hardware is limited in capacity. In addition, 
this type of cyber-physical system (CPS) heavily relies on real-time 
capabilities, making the scheduling precision a fundamental 
aspect to consider throughout the entire lifecycle. 

This study is mainly motivated by our experience with a self-driving 
truck depicted in \autoref{fig:Truck} that participated in the 2016 Grand Cooperative Driving 
Challenge (GCDC)\footnote{http://www.gcdc.net}. The competition includes 
a number of collaborative maneuvers that must be safely performed 
by participating vehicles. As there are several team members 
constantly working on the truck's software, it has become evident 
that the project would greatly benefit from a reliable, traceable, and structured deployment 
process. On the other hand, significant performance overhead
in the software execution must be avoided apparently.

\subsection{Research Goal \& Research Questions}

The goal of this work is to systematically evaluate the influence of sandboxed execution 
environments for applications from the automotive domain. 
We are particularly interested in studying the impact on two quality 
attributes of the system: Scheduling precision and input/output 
performance. Hence, in this article the following research questions 
are addressed:

\begin{description}
\item[\textit{RQ-1:}] \textit{How does the execution environment influence the scheduling precision of a given application?}
\item[\textit{RQ-2:}] \textit{How does the execution environment influence the input/output performance of a given application?}
\end{description}

\subsection{Contributions of the Article}

We demonstrate the impact of using container-based technology 
in relation to the underlying operating 
system (i.e., real-time kernel) on the performance of a given 
automotive application. To achieve this, we conduct a multi-method 
research approach that includes (i) an investigation of the current 
state-of-the-art in the area; (ii) a controlled experiment on desk 
using example applications; and (iii) a validation of the results in 
a real-world scenario using the aforementioned Volvo FH truck. 
To the best of our knowledge, this work is the first of its kind to 
evaluate the use of Docker in the autonomous vehicles business.

\subsection{Structure of the Article}

The remainder of this article is structured as follows. 
Sec.~\ref{sec:Methodology}, contains the methodology used in 
this work. In Sec.~\ref{sec:LiteratureReview}, the related work is 
presented in the form of a literature review. Sec.~\ref{sec:Experiments} 
contains a description of the experiments in a controlled environment and 
in the truck. In Sec.~\ref{sec:Results} we present the results, followed by 
an analysis and discussion in Sec.~\ref{sec:AnalysisDiscussion}. Finally, 
conclusion and future work are described in Sec.~\ref{sec:Conclusion}.

\section{Methodology} 
\label{sec:Methodology}

In order to achieve an understanding of how the execution environment 
influences the performance of applications, three studies are designed in 
a way that they complemented each other. First, we perform a literature 
review using key terms in the search, followed by the inclusion of additional 
papers using the snowballing approach \cite{Wohlin2014}. Then, we conduct 
a controlled experiment \textit{on desk} to measure the scheduling precision 
and I/O performance of sample applications when deployed on four different 
environments. Finally, we conduct an experiment using a real-world application 
that is currently deployed on the self-driving truck. The latter experiment is 
designed in a way that the findings from the first study can be validated. 

The goal with such design is to obtain meaningful results from different 
sources, combine them, and contribute towards a safe, robust, and reliable 
software deployment process in the domain of self-driving vehicles. 
Performing such multi-method research allows collection of data of different 
types, resulting in a wider coverage of the problem space \cite{Wood1999}.

\section{Literarature Review} 
\label{sec:LiteratureReview} 

Performing a literature review in a way that it is systematic brings a number of benefits. The steps are replicable, justified, and the results of the study provide basis for future research in a given area. In the present work, we perform a light-weight literature review based on the snowballing approach \cite{Wohlin2014}. The technique consists of searching for related work by investigating citations from a known set of relevant papers. In summary, the steps to conduct a snowballing procedure include (i) selecting a set of papers referred to as \textit{the start set}; (ii) apply forward snowballing (citation analysis) in the start set; and (iii) apply backward snowballing on each paper identified. This process iterates until no new relevant papers are found.

Our start set consists of papers found through a search in the Scopus digital library \cite{WWW_scopus}. The search string is presented in \autoref{tbl:searchstring}. It contains key terms concerning performance evaluation of virtualization/container approaches. We also include the term \textit{Docker} as authors may have different definitions for the framework, yet we are interested in capturing papers discussing it. The search resulted in 215 papers that had their titles and abstracts critically evaluated according to our inclusion and exclusion criteria. Inclusion criteria include papers presenting benchmarking activities and experiences with Docker. Exclusion criteria exclude, for example, papers in languages other than English and short papers pointing to online material.

The forward and backward snowballing procedures resulted in 10 highly relevant papers which were selected as primary studies. In addition to this set, we included 2 external papers that were not found during the search, even though they are relevant to the context of this study. The final set was critically appraised in the search for insights that would aid in the understanding of the state-of-the-art of software virtualization/containers. The next subsection contains selected outcomes from the search.

\begin{table}[h]
\footnotesize
\centering	
\caption{Search String in Scopus}

	\begin{tabular}{|c|}
	    \hline
		      TITLE-ABS-KEY(\textit{Performance} OR \textit{Comparison} OR \textit{Latency} \\
		      OR \textit{Evaluation} OR \textit{Container-Based} OR \textit{Linux Containers} \\
		      OR \textit{Lightweight Virtualization} OR \textit{Container Cloud} OR \textit{Docker}) \\
		      AND \\ 
		      (LIMIT-TO(SUBJAREA,``COMP'') 
	    	\\
	    \hline    
	\end{tabular} 

\label{tbl:searchstring}
\end{table}

\subsection{Outcomes of the Review}

The challenge and need for maintaining, deploying, and configuring software for
embedded, cyber-physical systems is identified in current literature
\cite{cberger,kryl}. Berger \cite{cberger} investigates a continuous deployment (CD) 
process for self-driving miniature vehicles using Docker containers. Multiple
containers are used to create binary packages from source that are signed and
tested for different hardware architectures, exemplifying a possible CD process
for CPSs.

The trade-off of virtualization is widely discussed in literature.
Krylovskiy \cite{kryl} evaluates Docker for resource constrained devices,
identifying negligible overhead and even outperforming native execution in some
tests. Similar performance characteristics are identified by Felter et al.~\cite{felt},
identifying negligible overhead for CPU and memory performance.
Felter et al. recommend that for input/output-intensive workloads, virtualization should
be used carefully. As Felter et al.~use the AUFS storage driver for their
evaluation, we use the Overlay storage driver as it is considered to be
the default option for Docker in the future.

Negligible performance overhead is also identified by Raho et al.~\cite{rago}
who identify a \SI{3.2}{\percent} CPU performance decay when using Docker on an ARM device.
In respect to self-driving vehicles, however, there lacks research using Docker
for real-time applications. Mao et al.~\cite{mao} investigate using Docker for
time-sensitive telecommunications networks in the cloud. A benchmarking tool is
used to report the computational worst case execution time when executed
natively and within a container using three different operating system kernels.
The results show that the difference between Docker and native execution is only
\SI{2}{\us} when using the preempt\_rt real-time Linux kernel irrespective to system
load. Latency when using the real-time kernel is improved by \num{13.9} times for
Docker (from \SI{446}{\us} to \SI{32}{\us}) and \num{7.8} times for native execution (from
\SI{234}{\us} to \SI{30}{\us}) in comparison to a vanilla kernel \cite{mao}. The authors
concluded that in order to satisfy real-time demands, a finely-tuned real-time
Linux kernel must be used to achieve near native performance.

Furthermore, the authors identified considerable overhead when using Docker
on multiple hosts. This finding is also confirmed by Ruiz et al.~\cite{ruiz} who identify
a high cost when using multiple containers on different nodes. The authors of
\cite{gonz} also investigate using Docker to realize a modular CPS architecture
design. The authors encapsulate computational nodes into containers to improve security and to
ease system administration through modularity and scalability, decoupling the
complexity of a CPS into smaller subsystems. They point out the benefit that
teams can work independently and concurrently on Docker images as well as
the need of using a real-time enabled Linux kernel. Current literature points to the
fact that Docker can be used for CPS due to negligible overhead identified in
current research. However, further evidence is needed for the area of automated driving.

\section{Experiments}
\label{sec:Experiments} 

The aim of the controlled experiment is to systematically evaluate the
scheduling precision and input/output performance of two sample applications;
both during native execution and encapsulated into a sandboxed environment
(Docker). Scheduling precision refers to how precisely, in measures of time, 
the CPU scheduler is able to execute an operation from when the operation was 
first called. Whereas the input/output performance refers to measuring the 
performance of camera input and disk output, namely the time it takes capturing 
an image and saving it on the disk. Through a sequence of controlled steps, 
the sample applications are executed in four different execution environments. 
The execution environments consist of an alternation of (i) executing the sample 
applications natively or sandboxed within a Docker container and (ii) executing 
the sample applications on a target system with a vanilla or a real-time enabled 
kernel. Understanding how the respective execution environments influence the 
scheduling precision and input/output performance will ultimately decide how 
deterministic, with respect to time, the system is to uncover the performance 
cost of using Docker for software deployment.

The two sample applications, named Pi and Pi/IO component, are realized with the
open-source development framework
OpenDaVINCI\footnote{http://www.opendavinci.org}. Measurement points in the form
of timestamps are captured during runtime of the sample applications to uncover
the timing behavior of the respective application. The Pi component, used to
measure scheduling precision, is tasked to calculate the next digit of Pi until
it reaches \SI{80}{\percent} of its designated time slice. The remaining \SI{20}{\percent} should be spent
sleeping the process. For the experiments, the \SI{80}{\percent} CPU load was established based on our observations 
when executing real-life scenarios with the truck. Four measurement points are captured during runtime: the
timing of OpenDaVINCI's middleware (named Overhead 1 and Overhead 2), the time
duration for calculating Pi (Pi Calculation) and the amount of time the process
sleeps (Sleep). The Pi/IO component, used to measure input/output performance,
is tasked to capture an image (Camera Input) and store it to disk (Disk Output).

Treatments used for assessing the impact of factors that are specific to the
execution environment (e.g. execution context and deployment context). During
execution of the sample applications, system load is applied to the target
system in order to traverse kernel code paths and to mimic run-time load of a self-driving 
vehicle. Load is applied to the system via a user-space
application (stress-ng\footnote{http://goo.gl/o5WuFW}), spawning two worker
threads that apply CPU load at \SI{80}{\percent} with scheduling priority 48. The controlled
experiment is executed on an AEC 6950 embedded personal computer \footnote{http://www.aaeon.com/en/p
/fanless-embedded-computers-aec-6950/}. The Linux
kernel version 3.18.25 is chosen for both vanilla and real-time enabled kernel.
A usb webcam Logitech c930e is used for measuring the I/O performance.
Ingo Molnar's real-time patch (preempt\_rt) is used to bring real-time
capabilities to the kernel. The Docker storage driver Overlay is used in the
experiment, instead of the default AUFS driver, which is known for performance
issues. 

The results from the controlled experiment give recommendation on which
execution environment is best suited to meet real-time requirements. That
respective execution environment is applied to the self-driving heavy vehicle
(Volvo FH16 truck), where parts of the controlled experiment are executed in
order to validate the results. The Pi Component is executed exclusively on the
self-driving truck as the camera hardware setup is networked and differ greatly 
from that of the controlled experiment and would require modifications 
to the sample application, which excludes the Pi/IO component for the uncontrolled tests.

The intention of the real-life use case is to further understand the impact of
using Docker for software deployment on a system with all required operations to
enable self-driving capabilities running simultaneously as system load. Consequently, the
environment is less-controlled from the researchers' access, however, it is more
realistic in the sense of operational load. The sample application Pi Component
is used to capture data, with the exact same execution parameters and data
collection procedure on precisely the same target system, where the difference between
the experiments is seen in the applied system stress.

\section{Results} 
\label{sec:Results}

\pgfplotstableread[col sep = semicolon]{./data/pi/colsum_load.csv}\mydatanoload
\pgfplotstableread[col sep = semicolon]{data/pi/colsd_load.csv}\mydataload
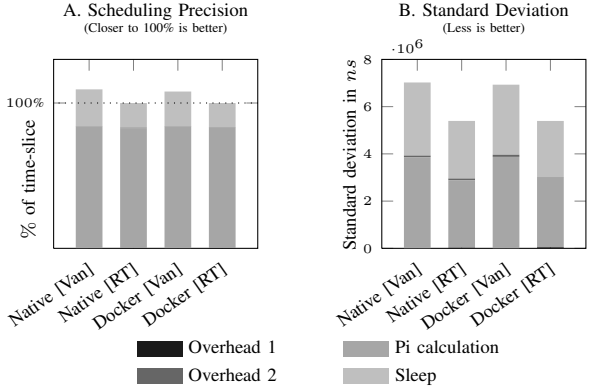
\begin{figure}[t]
\begin{tikzpicture}

\begin{axis}[
        title=\parbox{4cm}{\centering\scriptsize A. Scheduling Precision\\\tiny (Closer to 100\% is better)},
        name=mean,
        at={(0.3\linewidth,0)},
        anchor=west,
        ybar stacked,
        width=\textwidth*.24,
        height=\textheight*.17,
        ylabel={},
        xtick=data,
        ymin=0,
        ymax=1.3,
        enlarge x limits={abs=.8},
        ylabel={\scriptsize$\%$ of time-slice},
        xticklabels={\scriptsize Native [Van],\scriptsize Native [RT],\scriptsize Docker [Van],\scriptsize Docker [RT]},ytick={1,...,4},
        xticklabel style={anchor=north east,rotate=35},
        y label style={at={(axis description cs:-.04,0)},anchor=south west},
        yticklabel={\tiny\pgfmathparse{\tick*100}\pgfmathprintnumber{\pgfmathresult}\%},
        point meta={y*100}
        ] 

\draw[black, dotted] (axis cs:-2,1) -- (axis cs:11,1);
\addplot+[draw opacity=0,fill=orange,ybar,area legend] table[x=scenario,y=odv_oh1]{\mydatanoload};
\addplot+[draw opacity=0,fill=blues3,ybar,area legend] table[x=scenario,y=pi_calc]{\mydatanoload};
\addplot+[draw opacity=0,fill=blues5,ybar,area legend] table[x=scenario,y=odv_oh2]{\mydatanoload};
\addplot+[draw opacity=0,fill=blues1,ybar,area legend] table[x=scenario,y=sleep]{\mydatanoload};

\end{axis}

\begin{axis}[
        title=\parbox{2.4cm}{\centering\scriptsize B. Standard Deviation\\\tiny (Less is better)},
        name=std,
        at={(.8*\linewidth,0)},
        anchor=west,
        ybar stacked,
        width=\textwidth*.24,
        height=\textheight*.17,
        ylabel={},
        xtick=data,
        ticklabel style = {font=\tiny},
        ymin=0,
        ymax=8000000,
        enlarge x limits={abs=.8},
        ylabel={\scriptsize Standard deviation in $ns$},
        xticklabels={\scriptsize Native [Van],\scriptsize Native [RT],\scriptsize Docker [Van],\scriptsize Docker [RT]},xtick={1,...,4},
        xticklabel style={anchor=north east,rotate=35},
        y label style={at={(axis description cs:-.08,-0.06)},anchor=south west},
        point meta={y*100},
        legend style={
                draw=none, 
                text depth=0pt,
                at={(-.3,-0.43)},
                legend cell align=left,
                anchor=north,
                legend columns=2,
                column sep=.8cm,
                /tikz/every odd column/.append style={column sep=0cm},
            }
        ] 

\addplot+[draw opacity=0,fill=orange,ybar,area legend] table[x=scenario,y=odv_oh1]{\mydataload};
\addplot+[draw opacity=0,fill=blues3,ybar,area legend] table[x=scenario,y=pi_calc]{\mydataload};
\addplot+[draw opacity=0,fill=blues5,ybar,area legend] table[x=scenario,y=odv_oh2]{\mydataload};
\addplot+[draw opacity=0,fill=blues1,ybar,area legend] table[x=scenario,y=sleep]{\mydataload};

\legend{\scriptsize Overhead 1,\scriptsize Pi calculation,\scriptsize Overhead 2,\scriptsize Sleep};
\end{axis}
\end{tikzpicture}
\caption{Scheduling precision results from controlled experiment.}
\label{fig:pi-chart-load}
\end{figure}

In this section, the results from the experiments are introduced.
\autoref{fig:pi-chart-load} presents the results of running the first
experimental unit, namely Pi Component. It is run with a frequency of
\SI{100}{\Hz} in four execution environments. Both figures address the
scheduling precision of the execution environments. \autoref{fig:pi-chart-load}.A
depicts the average time deadline for each of the four execution environments,
while \autoref{fig:pi-chart-load}.B depicts how deterministic each of the
execution environments are, i.e.~how much does each executing time-slice
vary from the resulted mean time deadline.

Executing the experimental unit
on a system with a Linux vanilla kernel results in an average time deadline
violation of approximately \SI{10}{\percent}. \autoref{fig:pi-chart-load}.B presents the
most deterministic execution environment is executing the experimental
unit on a system with an preempt\_rt real-time Linux kernel. The standard
deviation of the sleep execution on a system with native execution and
Linux vanilla kernel is roughly \SI{2400}{\us}. The total standard deviation of
the same execution environment is approximately \SI{7000}{\us}. Both figures
show no noticeable difference between executing the experimental unit in
Docker or natively in respect to both the determinism or scheduling
precision of a system. Outliers, which are apparent, are included in the results and have not been disregarded.

\pgfplotstableread[col sep = semicolon]{./data/piio/colsum_load.csv}\mydatameanload
\pgfplotstableread[col sep = semicolon]{data/piio/colsd_load.csv}\mydatasdload
\begin{figure}[t]
\begin{tikzpicture}

\begin{axis}[
        title=\parbox{4cm}{\centering\scriptsize A. Input/Output Performance\\\tiny (Less is better)},
        name=mean,
        at={(0.3\linewidth,0)},
        anchor=west,
        ybar stacked,
        width=\textwidth*.24,
        height=\textheight*.17,
        ylabel={},
        ytick={.04,.08,.12},
        ymin=0,
        ymax=.14,
        enlarge x limits={abs=.8},
        ylabel={\scriptsize$\%$ of time-slice},
        xticklabels={\scriptsize Native [Van],\scriptsize Native [RT],\scriptsize Docker [Van],\scriptsize Docker [RT]},xtick={1,...,4},
        xticklabel style={anchor=north east,rotate=35},
        y label style={at={(axis description cs:-.2,.12)},anchor=south west},
        yticklabel={\tiny\pgfmathparse{\tick*100}\pgfmathprintnumber{\pgfmathresult}\%},
        point meta={y*100}
        ] 

\addplot+[draw opacity=0,fill=orange,ybar,area legend] table[x=scenario,y=camio]{\mydatameanload};
\addplot+[draw opacity=0,fill=blues3,ybar,area legend] table[x=scenario,y=diskio]{\mydatameanload};

\end{axis}

\begin{axis}[
        title=\parbox{2.4cm}{\centering\scriptsize B. Standard Deviation\\\tiny (Less is better)},
        name=std,
        at={(.8*\linewidth,0)},
        anchor=west,
        ybar stacked,
        width=\textwidth*.24,
        height=\textheight*.17,
        ylabel={},
        ticklabel style = {font=\tiny},
        ymin=0,
        enlarge x limits={abs=.8},
        ylabel={\scriptsize Standard deviation in $ns$},
        xticklabels={\scriptsize Native [Van],\scriptsize Native [RT],\scriptsize Docker [Van],\scriptsize Docker [RT]},xtick={1,...,4},
        xticklabel style={anchor=north east,rotate=35},
        y label style={at={(axis description cs:-.12,-0.06)},anchor=south west},
        point meta={y*100},
        legend style={
                draw=none, 
                text depth=0pt,
                at={(-.3,-0.43)},
                legend cell align=left,
                anchor=north,
                legend columns=2,
                column sep=.8cm,
                /tikz/every odd column/.append style={column sep=0cm},
            }
        ] 

\addplot+[draw opacity=0,fill=orange,ybar,area legend] table[x=scenario,y=camio]{\mydatasdload};
\addplot+[draw opacity=0,fill=blues3,ybar,area legend] table[x=scenario,y=diskio]{\mydatasdload};

\legend{\scriptsize Camera Input,\scriptsize Disk Output};
\end{axis}

\end{tikzpicture}
\caption{Input/Output performance results from controlled experiment.}
\label{fig:piio-chart-load}
\end{figure}
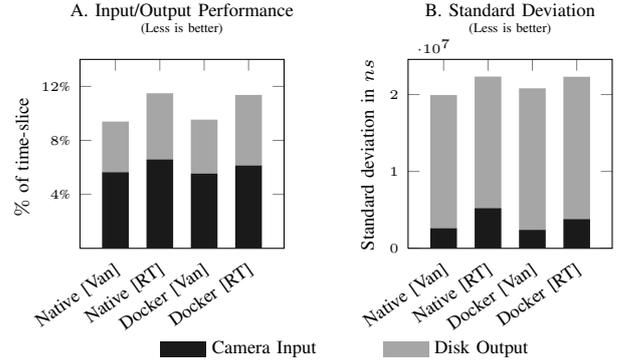

The two charts in \autoref{fig:piio-chart-load} present the camera and disk
performance for each of the execution environments running the second
experimental unit, namely Pi/IO Component, at \SI{10}{\Hz}. \autoref{fig:piio-chart-load}.A
shows that both operations of capturing an image and saving it to
disk consumes on average of approximately \SI{7}{\percent} and \SI{12}{\percent} of the total
time-slice. The execution environment with the worst input/output performance
is a system running with a preempt\_rt real-time Linux kernel and executing the experimental unit
either natively or in a Docker container. \autoref{fig:piio-chart-load}.B presents the standard deviation of the
input and output operations which depicts how deterministic each execution
environment is in regards to its input and output performance. Each of the
execution environments have approximately the same standard deviation.
\autoref{fig:piio-chart-load}.B shows that the input and output operations on
the system with an preempt\_rt real-time Linux kernel results in a higher
standard deviation, thus less deterministic in regards to such operations.

\begin{table}[t]
\scriptsize
\centering
\renewcommand{\arraystretch}{1.2}
\begin{tabu}{r|DDD}
\multicolumn{4}{c}{\hspace{3.5em}\textbf{Scheduling Precision}} \\
                                & \multicolumn{1}{c}{\small\textbf{$\eta^{2}$}} & \multicolumn{1}{c}{\scriptsize\textbf{Pillai's Trace}} & \multicolumn{1}{c}{\scriptsize\textbf{P-Value}}  \\
\scriptsize{deployment}         & 0.0001465922 & 0.000 & {$<$2.2e-16} \\
\scriptsize{kernel}             & 0.0677851134 & 0.068 & {$<$2.2e-16} \\
\scriptsize{deployment:kernel}  & 0.0001026772 & 0.000 & {$<$2.2e-16} \\
\multicolumn{4}{c}{\hspace{3.5em}\textbf{}}   \\
\multicolumn{4}{c}{\hspace{3.5em}\textbf{IO Performance}}   \\
                                & \multicolumn{1}{c}{\small\textbf{$\eta^{2}$}} & \multicolumn{1}{c}{\scriptsize\textbf{Pillai's Trace}} & \multicolumn{1}{c}{\scriptsize\textbf{P-Value}}  \\
\scriptsize{deployment}         & 0.0004096153 & 0.0004096 & {$<$2.2e-16} \\
\scriptsize{kernel}             & 0.0175100437 & 0.0175100 & {$<$2.2e-16} \\
\scriptsize{deployment:kernel}  & 0.0002103567 & 0.0002104 & {$<$2.2e-16}
\end{tabu}
\caption{effect size \& manova results}
\vspace{-3.5em}
\label{tbl:effect-pi}
\end{table}

A MANOVA test was conducted on all extracted data, including the outliers. This test
was conducted to understand the statistical impact each treatment have on the 
dependent variables, i.e. scheduling precision and input and output performance.
The MANOVA resulted in all treatments having a significant P-Value thus indicating a
significant impact for all treatments (i.e. alternating deployment and kernel). However,
the P-Value can not be fully trusted as this study has a vast amount of sample data which
carries a risk of Type I error \cite{kampenes2007systematic}, thus an effect size is 
extracted to fully comprehend what that significant impact is in reality.
\autoref{tbl:effect-pi} presents the results from the MANOVA as well as the effect sizes 
for each of the experimental units. The first three values depict the results from 
each of the alternating factors within the execution environment of the experimental 
unit Pi Component.

As the effect size and Pillai's Trace are both lower for the deployment treatment, in 
comparison with the kernel treatment, it suggests that switching between a vanilla 
Linux kernel and a preempt\_rt real-time Linux kernel has a greater impact on the 
scheduling precision and input/output performance in comparison to the alternation 
between executing natively and within a Docker container.

\pgfplotstableread[col sep = semicolon]{./data/truck/colsum_load.csv}\mydatameanload
\pgfplotstableread[col sep = semicolon]{data/truck/colsd_load.csv}\mydatasdload
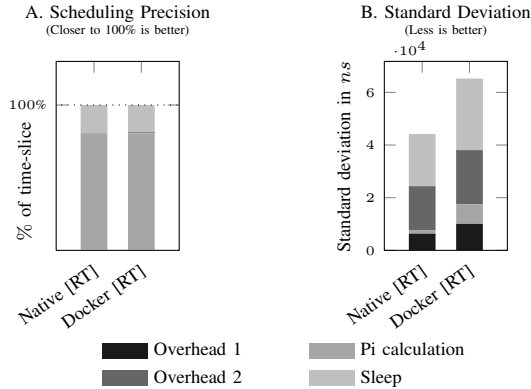
\begin{figure}[t]
\begin{tikzpicture}

\begin{axis}[
        title=\parbox{4cm}{\centering\scriptsize A. Scheduling Precision\\\tiny (Closer to 100\% is better)},
        name=mean,
        at={(0.36\linewidth,0)},
        anchor=west,
        ybar stacked,
        width=\textwidth*.18,
        height=\textheight*.17,
        ylabel={},
        xtick=data,
        ymin=0,
        ymax=1.3,
        enlarge x limits={abs=.8},
        ylabel={\scriptsize$\%$ of time-slice},
        xticklabels={\scriptsize Native [RT],\scriptsize Docker [RT]},ytick={1,2},
        xticklabel style={anchor=north east,rotate=35},
        y label style={at={(axis description cs:-.12,.0)},anchor=south west},
        yticklabel={\tiny\pgfmathparse{\tick*100}\pgfmathprintnumber{\pgfmathresult}\%},
        point meta={y*100}
        ] 

\draw[black, dotted] (axis cs:-2,1) -- (axis cs:11,1);
\addplot+[draw opacity=0,fill=orange,ybar,area legend] table[x=scenario,y=odv_oh1]{\mydatameanload};
\addplot+[draw opacity=0,fill=blues3,ybar,area legend] table[x=scenario,y=pi_calc]{\mydatameanload};
\addplot+[draw opacity=0,fill=blues5,ybar,area legend] table[x=scenario,y=odv_oh2]{\mydatameanload};
\addplot+[draw opacity=0,fill=blues1,ybar,area legend] table[x=scenario,y=sleep]{\mydatameanload};

\end{axis}

\begin{axis}[
        title=\parbox{2.4cm}{\centering\scriptsize B. Standard Deviation\\\tiny (Less is better)},
        name=std,
        at={(.86*\linewidth,0)},
        anchor=west,
        ybar stacked,
        width=\textwidth*.18,
        height=\textheight*.17,
        ylabel={},
        xtick=data,
        ticklabel style = {font=\tiny},
        ymin=0,
        enlarge x limits={abs=.8},
        ylabel={\scriptsize Standard deviation in $ns$},
        xticklabels={\scriptsize Native [RT],\scriptsize Docker [RT]},xtick={1,...,4},
        xticklabel style={anchor=north east,rotate=35},
        y label style={at={(axis description cs:-.2,-0.06)},anchor=south west},
        point meta={y*100},
        legend style={
                draw=none, 
                text depth=0pt,
                at={(-.8,-0.43)},
                legend cell align=left,
                anchor=north,
                legend columns=2,
                column sep=.8cm,
                /tikz/every odd column/.append style={column sep=0cm},
            }
        ] 

\addplot+[draw opacity=0,fill=orange,ybar,area legend] table[x=scenario,y=odv_oh1]{\mydatasdload};
\addplot+[draw opacity=0,fill=blues3,ybar,area legend] table[x=scenario,y=pi_calc]{\mydatasdload};
\addplot+[draw opacity=0,fill=blues5,ybar,area legend] table[x=scenario,y=odv_oh2]{\mydatasdload};
\addplot+[draw opacity=0,fill=blues1,ybar,area legend] table[x=scenario,y=sleep]{\mydatasdload};

\legend{\scriptsize Overhead 1,\scriptsize Pi calculation,\scriptsize Overhead 2,\scriptsize Sleep};
\end{axis}

\end{tikzpicture}
\caption{Scheduling precision results from the uncontrolled experiment.}
\vspace{-1.5em}
\label{fig:truck-chart-load}
\end{figure}

Finally, after executing the controlled experiment, an uncontrolled experiment
was carried out on a machine installed with the best suitable kernel found
through the controlled experiment, namely a preempt\_rt real-time Linux kernel.
\autoref{fig:truck-chart-load} presents the results from
executing the uncontrolled experiment on the self-driving truck with an applied
load produced by all software components used for its self-driving capabilities. 
The goal is to understand how Docker impacts the time critical
scheduling precision. \autoref{fig:truck-chart-load}.A shows that on average
neither a Docker nor a native execution of the experimental unit violated the
specified time deadline. \autoref{fig:truck-chart-load}.B suggests that a
deployment applying native execution performs better compared to executing
with a Docker container in regards to the determinism of all executions.
However, the standard deviation difference between the two deployment
approaches is a mere \SI{20}{\us}. The effect size ($\eta^{2}$) of the deployment
in the uncontrolled experiment is $0.5041$ which can be considered a
medium effect in regards to Cohen's D.

All experimental material such as raw data, experimental units, and
statistical R scripts used for extracting and processing the data presented in this
section can be found online.\footnote{https://github.com/docker-rt-research/experimental-material}

\section{Analysis \& Discussion} 
\label{sec:AnalysisDiscussion}

It is not sufficient to look at the impact of the deployment strategy
exclusively to fully comprehend what execution environment is suitable
to adopt for a system of a self-driving vehicle. Additional factors play
crucial parts of the performance within such a system. Further factors
such as kernel and system load are crucial to acknowledge when deciding
upon the execution environment. This is confirmed by the results, which
show that utilizing Docker for the deployment strategy of a self-driving
vehicle has negligible impact on the performance of the system.
The literature review further plays an integral role in the findings, 
as it has been pointed out in related work that a Docker solution does not 
add substantial overhead to a given system's performance.

The data gathered from all executions in both experiments have shown
that Docker is not the crucial factor to focus on when deciding which
execution environment to adopt for a self-driving vehicle. Both the
controlled and uncontrolled experiments presented similar results in
regards to the scheduling precision, whereas the controlled experiment
extended the scope for the input/output performance of the system.
The data extracted from each environment has shown that selecting
the correct kernel has a greater importance on the scheduling precision
and input/output performance of a self-driving vehicle, where both
effect size and the presented graphs convey convincing evidence.

This is in line with previous research exploring the impact of Docker
utilizing a preempt\_rt real-time Linux kernel (cf.~Mao et al.~\cite{mao}).
They present that the difference between a native execution and executing
within a Docker container is a mere \SI{2}{\us}. While the latency is improved
by $13.9$ times when utilizing a preempt\_rt real-time Linux kernel in
comparison to a generic Linux kernel. Further research has also shown
similar results when utilizing Docker for the deployment strategy where
Felter et al.~\cite{felt} found that Docker has negligible overhead in regards
to CPU and memory performance, and Krylovskiy \cite{kryl} presents negligible
overhead introduced by Docker when executed with an ARM CPU.

Other performance aspects such as input/output performance are taken
into consideration during the controlled experiment. The results show that
executing the application within a Docker container has negligible impact
on the input/output performance, while utilizing a preempt\_rt real-time
Linux kernel had a negative impact on the input/output performance of the
application. This may be explained by the system's preemptive approach
where the input/output operations can be preempted by other processes
and thus, increasing the time required to execute the operations.

The results regarding the determinism of the execution environment
captured in the uncontrolled experiment on the self-driving truck differ from
those captured in the controlled experiment. Where the uncontrolled
experiment shows that Docker has an impact on how deterministic a
system is when executed alongside components, which enable the
self-driving capabilities of the truck. However, the results suggest that
the load introduced in the controlled experiment is noticeably more
exhausting in relation to the truck experiment as its highest standard
deviation is around \SI{7000}{\us} while the truck's highest standard
deviation is around \SI{60}{\us}.

\subsection{Threats to Validity} The four perspective of validity threats
presented by Runeson and H\"{o}st \cite{runeson} are discussed. In respect to
construct validity, the sample applications are realized with OpenDaVINCI to
ensure a high degree of software correctness and completeness, meeting the
design requirements for real-time systems. For internal validity, a number of
strategies are used to limit the risk of an unknown factor impacting the data
collected. Namely, the execution of the sample applications is carried out by a
script to ensure precise reproducibility, all peripherals such as networking are
detached and data is collected via serial communication to limit additional load
to the system. In respect to external validity, the results of this study can be
applied to time-sensitive applications in respect to the hardware and software
used. The hardware used in this study is industrial grade, making the experiment
reproducible and the results relatable to similar contexts. For conclusion
validity, there exists a possibility of Type I and Type II statistical errors.
Due to the sample size of the data collected, Type I and Type II errors are
considerable since where an increasing sample size will result in a decreasing
P-value. For that reason, this study has put less emphasis on the P-value,
taking a larger consideration on the effect size when evaluating the data.

\section{Conclusion \& Future Work} 
\label{sec:Conclusion}

A literature review and two experiments, one controlled and one uncontrolled, have been
carried out to fully comprehend how an alternation of factors within
the execution environment influence the execution performance
of a system for a self-driving vehicle. More specifically, these experiments
sought to uncover which Linux kernel is most suitable for such a context,
and whether or not utilizing Docker as a technical platform for a software
deployment strategy has an impact on the time critical deadlines specified
for the real-time application.

Initially, the controlled experiment intended to identify the most appropriate
kernel in terms of scheduling performance. The most appropriate kernel was
later implemented into an uncontrolled environment of a self-driving truck,
which participated in the 2016 Grand Cooperative Driving Challenge (GCDC).
The research goal was to identify whether Docker is a suitable technical
environment to realize continuous integration, continuous deployment,
and continuous experimentation on the self-driving truck.
Our results show that Docker is not the critical factor to consider when
selecting an execution environment; however, the Linux kernel in use was
identified as having a greater impact on the scheduling precision and
input/output performance of the software. 

Future research is needed to understand how Docker behaves with respect to 
network performance, when several components are executed from within
Docker containers and there is communication between separate computer nodes. 
Moreover, on-road tests will be considered in future work in order to cover variables 
such as CPU load and memory footprint under real-life conditions.

\bibliographystyle{IEEEtran} 
\bibliography{references}

\begin{thebibliography}{10}
\providecommand{\url}[1]{#1}
\csname url@rmstyle\endcsname
\providecommand{\newblock}{\relax}
\providecommand{\bibinfo}[2]{#2}
\providecommand\BIBentrySTDinterwordspacing{\spaceskip=0pt\relax}
\providecommand\BIBentryALTinterwordstretchfactor{4}
\providecommand\BIBentryALTinterwordspacing{\spaceskip=\fontdimen2\font plus
\BIBentryALTinterwordstretchfactor\fontdimen3\font minus
  \fontdimen4\font\relax}
\providecommand\BIBforeignlanguage[2]{{%
\expandafter\ifx\csname l@#1\endcsname\relax
\typeout{** WARNING: IEEEtran.bst: No hyphenation pattern has been}%
\typeout{** loaded for the language `#1'. Using the pattern for}%
\typeout{** the default language instead.}%
\else
\language=\csname l@#1\endcsname
\fi
#2}}

\bibitem{Savor2016}
\BIBentryALTinterwordspacing
T.~Savor, M.~Douglas, M.~Gentili, L.~Williams, K.~Beck, and M.~Stumm,
  ``Continuous deployment at facebook and oanda,'' in \emph{Proceedings of the
  38th International Conference on Software Engineering Companion}, ser. ICSE
  '16.\hskip 1em plus 0.5em minus 0.4em\relax New York, NY, USA: ACM, 2016, pp.
  21--30. [Online]. Available: \url{http://doi.acm.org/10.1145/2889160.2889223}
\BIBentrySTDinterwordspacing

\bibitem{WWW_docker}
\BIBentryALTinterwordspacing
What is docker? [Online]. Available: \url{http://www.docker.com/what-docker}
\BIBentrySTDinterwordspacing

\bibitem{Mer14}
\BIBentryALTinterwordspacing
D.~Merkel, ``{Docker: lightweight Linux containers for consistent development
  and deployment},'' \emph{Linux Journal}, no. 239, Mar. 2014. [Online].
  Available: \url{http://dl.acm.org/citation.cfm?id=2600241}
\BIBentrySTDinterwordspacing

\bibitem{WWW_lmctfy}
\BIBentryALTinterwordspacing
Let me contain that for you (lmctfy). [Online]. Available:
  \url{https://github.com/google/lmctfy}
\BIBentrySTDinterwordspacing

\bibitem{masek}
P.~Masek and M.~Thulin, ``Container based virtualisation for software
  deployment in self-driving vehicles,'' Master Thesis, Chalmers University of
  Technology, 2016.

\bibitem{Wohlin2014}
\BIBentryALTinterwordspacing
C.~Wohlin, ``Guidelines for snowballing in systematic literature studies and a
  replication in software engineering,'' in \emph{Proceedings of the 18th
  International Conference on Evaluation and Assessment in Software
  Engineering}, ser. EASE '14.\hskip 1em plus 0.5em minus 0.4em\relax New York,
  NY, USA: ACM, 2014, pp. 38:1--38:10. [Online]. Available:
  \url{http://doi.acm.org/10.1145/2601248.2601268}
\BIBentrySTDinterwordspacing

\bibitem{Wood1999}
\BIBentryALTinterwordspacing
M.~Wood, J.~Daly, J.~Miller, and M.~Roper, ``Multi-method research: An
  empirical investigation of object-oriented technology,'' \emph{Journal of
  Systems and Software}, vol.~48, no.~1, pp. 13 -- 26, 1999. [Online].
  Available:
  \url{http://www.sciencedirect.com/science/article/pii/S0164121299000424}
\BIBentrySTDinterwordspacing

\bibitem{WWW_scopus}
\BIBentryALTinterwordspacing
Scopus digital library. [Online]. Available: \url{http://www.scopus.com}
\BIBentrySTDinterwordspacing

\bibitem{cberger}
C.~Berger, ``Testing continuous deployment with lightweight multi-platform
  throw-away containers,'' \emph{41th Conference on Software Engineering and
  Advanced Applications (SEAA)}, 2015.

\bibitem{kryl}
A.~Krylovskiy, ``Internet of things gateways meet linux containers: Performance
  evaluation and discussion,'' in \emph{Internet of Things (WF-IoT), 2015 IEEE
  2nd World Forum on}, Dec 2015, pp. 222--227.

\bibitem{felt}
W.~Felter, A.~Ferreira, R.~Rajamony, and J.~Rubio, ``An updated performance
  comparison of virtual machines and linux containers,'' in \emph{Performance
  Analysis of Systems and Software (ISPASS), 2015 IEEE International Symposium
  on}, March 2015, pp. 171--172.

\bibitem{rago}
M.~Raho, A.~Spyridakis, M.~Paolino, and D.~Raho, ``Kvm, xen and docker: A
  performance analysis for arm based nfv and cloud computing,'' in
  \emph{Information, Electronic and Electrical Engineering (AIEEE), 2015 IEEE
  3rd Workshop on Advances in}, Nov 2015, pp. 1--8.

\bibitem{mao}
C.~N. Mao, M.~H. Huang, S.~Padhy, S.~T. Wang, W.~C. Chung, Y.~C. Chung, and
  C.~H. Hsu, ``Minimizing latency of real-time container cloud for software
  radio access networks,'' in \emph{2015 IEEE 7th International Conference on
  Cloud Computing Technology and Science (CloudCom)}, Nov 2015, pp. 611--616.

\bibitem{ruiz}
\BIBentryALTinterwordspacing
C.~Ruiz, E.~Jeanvoine, and L.~Nussbaum, ``Performance evaluation of containers
  for hpc,'' in \emph{Euro-Par 2015: Parallel Processing Workshops: Euro-Par
  2015 International Workshops, Vienna, Austria, August 24-25, 2015, Revised
  Selected Papers}.\hskip 1em plus 0.5em minus 0.4em\relax Cham: Springer
  International Publishing, 2015, pp. 813--824. [Online]. Available:
  \url{http://dx.doi.org/10.1007/978-3-319-27308-2_65}
\BIBentrySTDinterwordspacing

\bibitem{gonz}
\BIBentryALTinterwordspacing
P.~Gonz{\'a}lez-Nalda, I.~Etxeberria-Agiriano, I.~Calvo, and M.~C. Otero, ``A
  modular cps architecture design based on ros and docker,''
  \emph{International Journal on Interactive Design and Manufacturing
  (IJIDeM)}, pp. 1--7, 2016. [Online]. Available:
  \url{http://dx.doi.org/10.1007/s12008-016-0313-8}
\BIBentrySTDinterwordspacing

\bibitem{kampenes2007systematic}
V.~B. Kampenes, T.~Dyb{\aa}, J.~E. Hannay, and D.~I. Sj{\o}berg, ``A systematic
  review of effect size in software engineering experiments,''
  \emph{Information and Software Technology}, vol.~49, no.~11, pp. 1073--1086,
  2007.

\bibitem{runeson}
\BIBentryALTinterwordspacing
P.~Runeson and M.~H\"{o}st, ``Guidelines for conducting and reporting case
  study research in software engineering,'' \emph{Empirical Softw. Engg.},
  vol.~14, no.~2, pp. 131--164, Apr 2009. [Online]. Available:
  \url{http://dx.doi.org/10.1007/s10664-008-9102-8}
\BIBentrySTDinterwordspacing

\end{thebibliography}

\end{document}